\RequirePackage{fix-cm}
\documentclass[twocolumn]{svjour3}           
\smartqed  
\usepackage{graphicx}
\usepackage{amsmath}
\usepackage{subfig}
\usepackage{xcolor}
\usepackage{hyperref}
\usepackage{flushend}

\usepackage{microtype}  
\usepackage{float}      

\usepackage[section]{placeins}
\graphicspath{ {images/} }
\begin{document}

\title{Can Learned Frame-Prediction Compete with Block-Motion Compensation for Video Coding?
\thanks{A. M. Tekalp acknowledges support from the TUBITAK project 217E033 and Turkish Academy of Sciences (TUBA).}
}

\titlerunning{Can Learned Frame-Prediction Compete with Block-Motion Compensation}        

\author{Serkan Sulun  \and  A. Murat Tekalp 
}


\institute{College of Engineering, Koç University, 34450 Istanbul, Turkey
              \email{mtekalp@ku.edu.tr}           
}

\date{Received: 25 December 2019 / Accepted: 16 July 2020}

\maketitle

\begin{abstract}
Given recent advances in learned video prediction, we investigate whether a simple video codec using a pre-trained deep model for next frame~prediction based on previously encoded/decoded frames without sending any motion side information can compete with standard video codecs based on block-motion compensation. Frame differences given learned frame predictions are encoded by a standard still-image (intra) codec. Experimental results show that the~rate-distortion performance of the simple codec with symmetric complexity is on average better than that of x264 codec on 10~MPEG test videos, but does not yet reach the level of x265 codec. This result demonstrates the power of learned frame prediction (LFP), since unlike motion compensation, LFP does not use information from the current picture. The implications of training with $\ell^1$, $\ell^2$, or combined $\ell^2$ and adversarial loss on prediction performance and compression efficiency are analyzed.
\end{abstract}
\vspace{-12pt}

\section{Introduction}
\label{intro}
\vspace{-6pt}
An essential component of video compression is \textit{motion compensation}, which reconstructs a predicted frame with the help of block-based motion vectors sent as side information. Naturally this prediction is imperfect, so the \textit{residual frame difference} needs to be encoded and transmitted alongside \textit{motion vectors}. These two components constitute a compressed video file, whose size can be adjusted by rate-distortion (RD) optimization creating videos with varying bitrate and visual quality.
 
Until recently, there was no serious competition to block motion compensation (BMC) to form a predicted video frame. Advances in network architectures, training methods, and the graphical processing units (GPU) have enabled creation of powerful learned models for many tasks, including prediction of future video frames given past frames without using motion vectors. 

This paper investigates whether learned frame prediction (LFP) can replace the traditional BMC in video compression, making estimating and sending motion vectors as side information redundant. LFP is not constrained by the block translation motion model, but only uses previously decoded frames at both the encoder and decoder, unlike traditional BMC, which has access to the current frame at the encoder. On the other hand, a video codec employing LFP can use bits saved by not sending motion vectors to better code the prediction residual in order to increase video fidelity. Hence, it is of interest to analyze how a video codec based on LFP compares with traditional codecs.


The main contribution of this paper is to demonstrate that a simple video encoder based on a pre-trained LFP model yields rate-distortion performance that on the average exceeds that of the well established x264 encoder in sequential configuration. More generally, we provide answers to the following questions: \vspace{-6pt}
\begin{itemize}
\item How do we evaluate the performance of LFP models in video compression vs. computer vision? 
\item Can LFP compete with block-motion compensation in terms of compression rate-distortion performance? 
\item Does training with sum of absolute differences ($\ell^1$) or mean square ($\ell^2$) loss or a weighted combination of $\ell^2$ and adversarial losses provide better rate-distortion performance in video compression? 
\end{itemize}
\vspace{-3pt}

In the following, Section~\ref{related} reviews related works. Learned video frame prediction is discussed in Section~\ref{framepred}, and compression of the predictive frame differences is described in Section~\ref{videocomp}. Experimental results are presented in Section~\ref{results}. Section~\ref{conclusion} concludes~the~paper.
\vspace{-10pt}


\section{Related Work}
\label{related}
\vspace{-6pt}
A recent review of video prediction methods using neural networks can be found in~\cite{survey}.
Different from~\cite{survey}, we classify prior work on LFP in terms of the prediction methodology they employ and the loss function they use in training. 
In terms of prediction methodology employed, we classify LFP methods as \textit{frame reconstruction} methods that directly predict pixel values, and \textit{frame transformation} methods that predict transformation parameters, e.g., affine parameters, to transform the current frame into future frame. 

Among frame reconstruction methods, 
Srivastava {\it et al.}~\cite{srivastava} use long short-term memory (LSTM) autoencoders to simultaneously reconstruct the current frame and predict a future frame. 
Mathieu {\it et al.}~\cite{mathieu} propose multi-scale generator and discriminator networks and introduce a new loss that calculates the difference between gradients of predicted and ground-truth images. 
Kalchbrenner {\it et al.}~\cite{kalchbrenner} introduces an encoder-decoder architecture called Video Pixel Networks, where the decoder employs a PixelCNN~\cite{pixelcnn} network. 
Denton {\it et al.}~\cite{denton2018stochastic} propose a variational encoder to estimate a probability distribution for potential  predicted frame outcomes.  

Among frame transformation methods, 
Amersfoort {\it et al.}~\cite{amersfoort} predict affine transformation between patches from consecutive frames, which is applied on the current frame to generate the next one. 
Vondrick {\it et al.}~\cite{vondrick} also predict frame transformations but train the model using adversarial loss only. 
Villegas {\it et al.}~\cite{villegaslongterm} focus on long term prediction of human actions using pose information from a pretrained LSTM autoencoder as input.  
In a follow up work, Wichers {\it et al.}~\cite{wichers2018hierarchical} replace the pre-trained pose extracting network with a trainable encoder, enabling end-to-end training. 
Finn {\it et al.}~\cite{finn} predict object motion kernels using convolutional LSTMs. In a follow-up work, Babaeizadeh {\it et al.}~\cite{babaeizadeh2017stochastic} supplement Finn's model with a variational encoder to avoid generating blurry predictions. 
In a further follow-up, Lee {\it et al.}~\cite{lee2018stochastic} use adversarial loss to get more realistic results.

In terms of loss functions, most methods use mean square ($\ell^2$) loss.  Other loss functions used include $\ell^1$ loss \cite{mathieu} and cross-entropy loss \cite{srivastava,kalchbrenner}. Mathieu {\it et al.}~\cite{mathieu} introduce the gradient loss. Variational autoencoders employ KL-divergence \cite{mathieu}, \cite{vondrick},  \cite{villegaslongterm}, \cite{lee2018stochastic} and GANs employ adversarial loss~\cite{babaeizadeh2017stochastic},~\cite{lee2018stochastic},~\cite{denton2018stochastic}. Perceptual loss is also used, by comparing frames at a feature space~\cite{perceptualloss}, using a pretrained network \cite{villegaslongterm}. 

\hspace{-8pt} Our paper differs from other video prediction work~as: \vspace{-16pt} 
\begin{itemize}
\item While most related works use some form of LSTM, we chose a deep convolutional residual network, insired by EDSR~\cite{edsr}, for frame prediction. The rationale for this choice is explained in Section 3.
\item  In applications where the predicted frame is the final product, the visual quality, i.e., textureness and sharpness, of the predicted image is important; hence, the use of adversarial loss is well justified. However, most methods using such loss function do not report a direct quantitative evaluation of generated vs. ground-truth images using peak signal-to-noise ratio (PSNR) or structural similarity metric (SSIM). In~contrast, in video compression, it is customary to compare compression efficiency by the rate-distortion (RD) performance, where distortion is measured in terms of PSNR and predicted frames are only intermediate results, not to be viewed. Our results show that training with only $\ell^1$ or $\ell^2$ loss provides the~best RD  performance even though predicted frames may look blurry. 
\end{itemize}
\vspace{-3pt}

The state of the art video compression standard is high efficiency video coding, known as H.265/HEVC~\cite{heiko2016}. Several works to enhance H.265/HEVC codecs with or without deep learning have been proposed~\cite{lu2018}. While there are many works on learned intra-prediction or learned end-to-end image/video compression (e.g., \cite{dumas2018}), few works address learned frame prediction for video compression. Our work differs from them as follows: \vspace{-4pt}
\begin{itemize} 
\item  Several researchers propose learned models that supplement standard block motion compensation to enhance prediction and improve the compression efficiency of HEVC~\cite{iscas2018,lin2018,wang2018,zhao2018,xia2019,cho2019}. In contrast, we do not use block motion compensation at all. 
\item Chen {\it et al.}~\cite{zchen2018} employ a neural network for frame prediction. However, they also estimate and use 4x4 block motion vectors both at the encoder and decoder, even though they don't transmit them. In contrast, we do not need motion vectors at all. 
\end{itemize}
\vspace{-12pt}


\section{Learned Video Frame Prediction}
\label{framepred}
\vspace{-6pt}
Recurrent models, such as LSTM, has been the top choice of architecture to solve sequence learning problems. With the advent of ResNet, which introduces skip connections, it has become easy to train deep CNN to learn temporal dynamics from a fixed number of past frames. Although, in theory LSTMs can remember the entire history of a video, in practice due to training using truncated backpropagation through time~\cite{bai2018}, we obtain as good if not better performance by processing only a fixed number of past frames using a CNN.
Our approach is consistent with a recent study, where Villegas et al.~\cite{nips2019} show that large scale networks can perform state of the art video prediction without using optical flow or other other inductive bias in the network architecture.
In the following, we present our generator network architecture in Section~\ref{generator} and discuss the~details of training procedures in Section~\ref{training}.
\vspace{-10pt}

\subsection{The Generator Network} 
\label{generator}
\vspace{-6pt}
The architecture of our LFP network is inspired by the success of the enhanced deep super-resolution (EDSR) network~\cite{edsr} for single-image super-resolution (SISR). EDSR won the NTIRE 2017 Challenge on SISR~\cite{ntire} and a variation of it won NTIRE 2018 Challenge~\cite{ntire2018}. 

\begin{figure}[b]
\centering
	\includegraphics[scale=0.25]{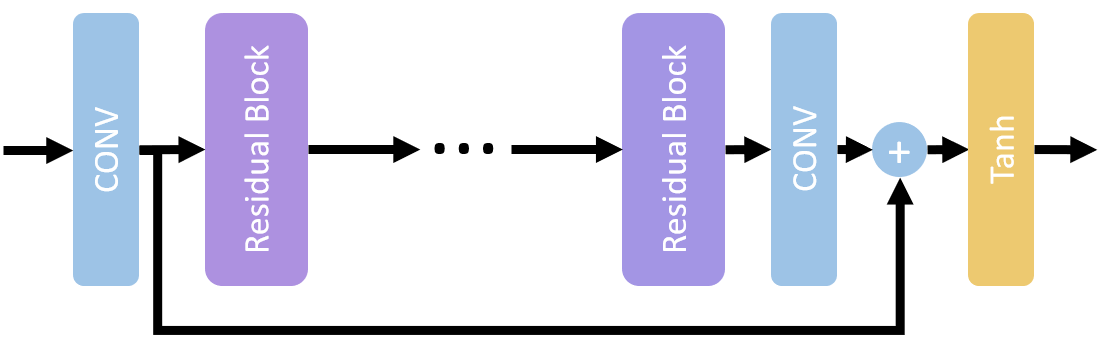} \vspace{-3pt} \\
	(a) \vspace{8pt}\\
	\includegraphics[scale=0.2]{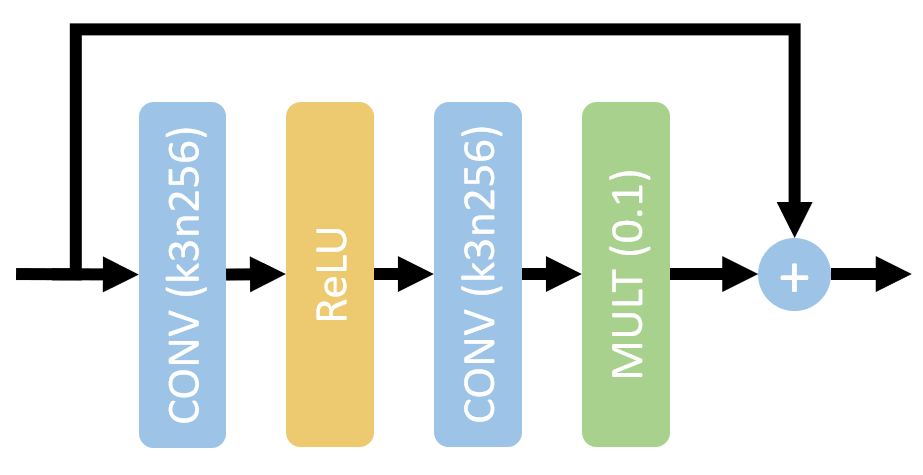} \vspace{-2pt} \\
	(b) \vspace{-4pt}\\
\caption{Block diagram of LFP-EDSR network: (a)~EDSR network with modified input and output layers, (b) each residual block (purple) of EDSR network.}
\label{fig:resnet}
\vspace{-3pt}
\end{figure}

In contrast to the original EDSR that takes a single low resolution color image as input, our LFP-EDSR network takes $K$ past greyscale frames as input and outputs a single greyscale frame. Since the~input and output frames are the same size, we don't need an upscaling layer \cite{pixelshuffler}. To this effect, we modified both the input and output layers of the EDSR network. The~resulting early fusion LFP-EDSR, whose architecture is depicted in Figure~\ref{fig:resnet}, is convolutional; hence, it can process video with arbitrary size (height and width). Only  the input layer processes groups of $K$ frames; hence, the~additional cost of processing $K$ input frames (instead of a single frame) is insignificant compared to~the overall computational cost of the network. As demonstrated in Section~\ref{results}, the~performance of the modified EDSR network to predict frames is surprisingly good. 

We use 32 residual blocks, where the convolution kernel is 3$\times$3 with padding 1 and the channel depth is 256 for all blocks. Rectified linear unit (ReLU) was used as the activation function for all hidden layers. The height and width of all intermediate layer activations and the output layer are the same as those of the input layer. After each residual block, residual scaling with 0.1 is applied~\cite{residualscaling}. At the output layer, output values are scaled between -1 and 1 during training. At test time, the~outputs are scaled between 0 and 255. 
\vspace{-8pt}

\subsection{Training}
\label{training}
\vspace{-10pt}
An important factor that affects video prediction and compression performance is the choice of loss function, which is related to how we evaluate the prediction performance of the network. Stochastic sampling methods, such as variational autoencoders and generative adversarial networks (GAN), produce sharper predicted images with more spatial details. However, they do so at the expense of higher mean square error~(MSE), which implies less fidelity to the ground-truth. In video compression the predicted image is only an intermediate result (not to be viewed), and the goal is to maximize the fidelity of predicted frame in order to minimize the bitrate to send the prediction error. In this context, training to minimize MSE yields better RD performance even if the predicted images may look blurry.

While minimization of the MSE implies maximization of the PSNR for videos in the training set, it is important to see how well this generalizes to videos in the test set.~It~was observed that models trained by $\ell^1$~loss generalize better~\cite{mathieu}, since minimization of $\ell^1$~loss converges to median prediction, whereas minimization of $\ell^2$ loss converges to mean prediction. 
\vspace{-2pt}

\subsubsection{Training Dataset}
\vspace{-6pt}
Our training dataset consists of two~million gray-level patch sequences, extracted from the UCF101 dataset, which contains 13K videos with 101 actions in 5 action types \cite{ucf101}. Each  patch sequence is $48 \times 48$ pixels with 9 frames. For each sequence, we predict the~9th frame given the first 8, and use the 9th frame as ground-truth. 

Regarding the length of training patch sequences, we observed a steady increase in PSNR performance as we used 2, 3, 4, 5, and 6 past frames to predict the next frame in our experiments. The PSNR performance leveled at using 6 past frames to predict the 7th frame. To better utilize GPU capabilities, we used 8 past frames, which is a power of 2, following the practice of choosing minibatch size \cite{chollet}. Our use of past 8 frames to predict the 9th is consistent with other frame prediction literature \cite{survey}. We selected a patch size of $48\times48$ following the original use of EDSR network for super-resolution \cite{edsr}. We also experimented with using a patch size of $96\times96$, which provided minimal PSNR improvements that is not worthy of the added computational complexity. We observed that $48\times48$ patches are sufficient to capture the motion well, considering the frame sizes in UCF101 dataset is $320\times240$. We note that use of $48\times48$ patches is consistent with other frame prediction literature \cite{survey}.

In extracting patch sequences, we select the video, the starting frame, and patch location on the start frame randomly. An extracted patch sequence is accepted if it contains sufficient motion, i.e., the mean square difference between successive pairs of frames exceeds a threshold. Patch sequences that do not satisfy the condition are accepted with probability 0.05. 
\vspace{-4pt}


\subsubsection{Training with $\ell^p$ Loss}
\vspace{-6pt}
We first trained our model based only on $\ell^p$ loss, where $p=1$ or $p=2$. 
We compute $\ell^p$ loss, given by
\begin{equation} \label{mse}
\ell^p = \frac{1}{N} \sum_{i=1}^{N} |y_i - x_i|^p
\end{equation}
over the 9th patch only, where $x$ and $y$ denote the generated and ground-truth 9th patch, respectively, and $i$~is the index looping over all $N$ pixels in a patch.

We used Adam optimizer \cite{adam} with an initial learning rate 1e-4 and a batch size 32. We trained our~model for 400,000 iterations, which lasted about 8 days using an NVIDIA GeForce GTX 1080Ti GPU on a HP Server with Intel Xeon Gold CPU @2.30GHz and 24 cores. 

\subsubsection{Training with $\ell^2$ and Adversarial Losses}
\begin{figure}[b]
	\centering
	\includegraphics[scale=0.30]{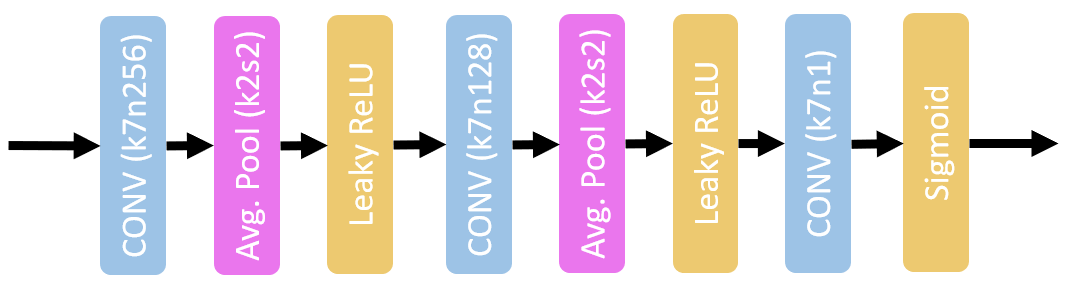}
	\caption{Block diagram of the discriminator network.}
	\label{fig:discriminator}
\vspace{-2pt}
\end{figure}

\vspace{-6pt}
We next train with combined $\ell^2$ and adversarial loss that requires a discriminator network. 
The block diagram of discriminator network is depicted in Figure~\ref{fig:discriminator}. There are three convolutional layers with kernel size 7$\times$7 and varying channel depths. Even though the discriminator is fully convolutional, by using pooling~layers with kernel size 2$\times$2 and stride 2, and avoiding padding, we obtain a scalar output when input size is 48$\times$48 pixels. For stable training, we used average pooling instead of max pooling to avoid sparse gradients as recommended~by~\cite{ganhacks}. Leaky rectified linear unit (Leaky ReLU) with a slope of 0.2 is used as the nonlinearity for the hidden layers as advised by~\cite{dcgan}. The final nonlinearity is a sigmoid, enabling an output value between 0 and 1, regarded as the probability that the input image is real (higher means more realistic). Note that the~discriminator network is not needed at test time.

We jointly trained randomly initialized discriminator and pretrained (by $\ell^2$ only) generator networks. Minibatch sizes are 16 for the generator and 32 for the discriminator. A minibatch for training the discriminator network consists of 16 ground-truth (real) and 16 generated (fake) patch-sequence samples. Generated samples are composed of the first 8 original patches concatenated with the predicted 9th patch that is the output of the generator. By feeding sequences instead of single patches into the discriminator, the discriminator can evaluate temporal continuity of motion similar to~\cite{mathieu}. 
We use target labels of 0 and 1 to indicate generated and original samples, respectively. Then, the binary cross entropy loss per sample is given by
\begin{equation}
L^D_{BCE} = -y\log x - (1-y)\log (1-x)
\label{eqn:two}
\end{equation}
where $x$ is the output score of the discriminator network (between 0 and 1) and $y$ is the binary ground-truth label. Note that if $y=0$ the first term, or else $y=1$ the second term is zero.  The discriminator is trained by accumulating the loss given by (\ref{eqn:two}) over a minibatch.

The loss function for the generator network is a combination of mean square loss and adversarial loss per generated sample with weights $\lambda_{MS}$ and $\lambda_{ADV}$, respectively, similar to \cite{mathieu}. We calculate the adversarial loss per generated sample by feeding each generated sample to the discriminator network and use a target label of 1, in order to quantify how far away each generated sample is from fooling the discriminator. Thus, the~adversarial loss per generated sample is defined as
\begin{equation}
L^G_{ADV} = -\log x_{disc}
\end{equation}
and the combined loss per sample for the generator is
\begin{equation}
L^G = \lambda_{MS} \left(\frac{1}{N} \sum_{i=1}^{N}(y_i - x_i)^2 \right) - \lambda_{ADV} \log x_{disc}
\label{eqn:comb}
\end{equation}
where $x_i$ denotes each pixel in generator's output, $x_{disc}$ is discriminator's output, $N$ is the number of pixels, $\lambda_{MS}=0.95$ and $\lambda_{ADV}=0.05$ are weights for the mean square and adversarial losses, respectively. The learning rates are constant at 1e-6 and 1e-5 for the generator and discriminator, respectively. The generator network is trained by accumulating the loss (\ref{eqn:comb}) over a minibatch. The adversarial training for 300,000 steps lasted 5 days. 

The inclusion of adversarial loss increases sharpness of predicted frames at the expense of higher MSE as shown in Sec.~\ref{res:pred}. Hence, choosing $\lambda_{ADV}$ and $\lambda_{MS}$ is a matter of trade-off between sharper looking images and lower MSE. This is consistent with observations in the SISR literature that increasing $\lambda_{ADV}$ increases the output sharpness at the expense of higher MSE~\cite{esrgan}. The values $\lambda_{MS}=0.95$ and $\lambda_{ADV}=0.05$, offer a good balance between sharpness and MSE~\cite{mathieu}.
\vspace{-10pt}


\section{Compression of Predictive Frame Differences}
\label{videocomp}
\vspace{-6pt}

Since the LFP model is sent to the decoder only once, the video codec is free from motion vector overhead.
\vspace{-12pt}

\subsection{Encoder}
\begin{figure}[t]
	\centering
	\includegraphics[scale=0.31]{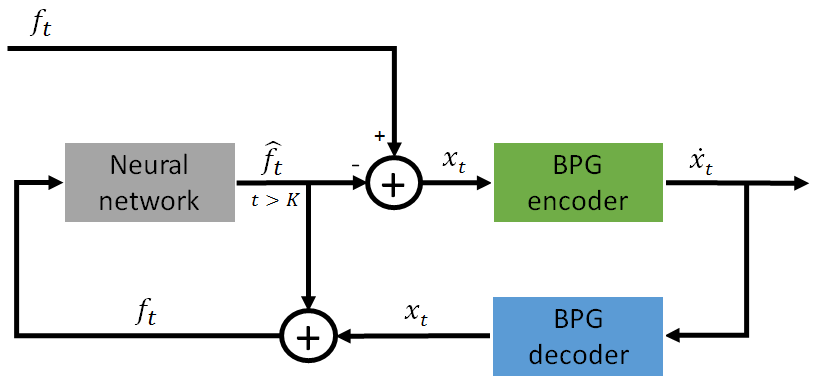} \\
	(a) \vspace{6pt} \\
	\includegraphics[scale=0.33]{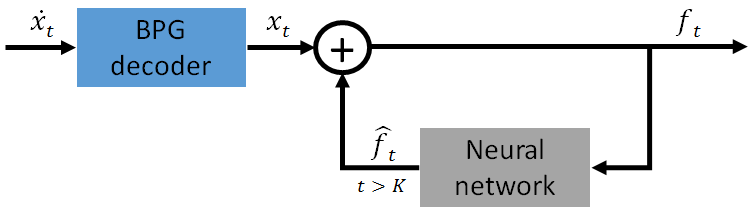} \\
	(b) \vspace{-4pt}
	\caption{Block diagram of video codec using LFP and BPG still-image codec: (a) Encoder, (b) Decoder.}
\label{fig:codec}
\end{figure}

\vspace{-6pt}
The block diagram of the proposed video encoder is depicted in Figure~\ref{fig:codec}(a).
The first $K$ frames are input to the Better Portable Graphics (BPG) encoder~\cite{bpg} without prediction (encoded as I pictures). The~neural network starts predicting next frame with frame $K + 1$ given previous $K$ decoded frames. In order to input exactly the same past frames into the~neural network (LFP model) in the encoder and decoder, the encoder has a BPG decoder in the feedback loop, which reproduces decoded frames. The decoded differences are added to the next frame predictions in order to produce decoded frames that are identical to the ones at the~decoder, which become the input to the~neural network at the~next time step. 
\vspace{-10pt}

\subsection{Decoder}
\vspace{-6pt}
The decoder also runs a neural network with the same model parameters to predict the next frame given the previous $K$~decoded frames. 
The first $K$ frames are received as intra-coded frames. The block diagram of the proposed video decoder is presented in Figure~\ref{fig:codec}(b).
\vspace{-12pt}


\section{Evaluation and Results}
\label{results}
\vspace{-6pt}
Our test dataset consists of 10 MPEG test sequences in grayscale format. We analyze frame prediction results in Section~\ref{res:pred}. Section~\ref{res:codec} evaluates the compression efficiency of codec using LFP+BPG vs. standard codecs.

\begin{figure}[b]
\centering
    \subfloat[Predicted frame obtained by LFP-L2 model  ]{\includegraphics[width=.99\linewidth]{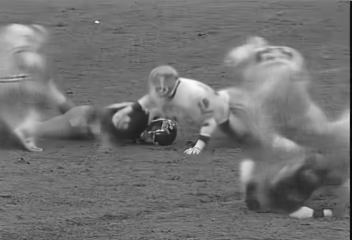}} \vspace{-6pt} \\
    \subfloat[Predicted frame obtained by LFP-GAN model   ]{\includegraphics[width=.99\linewidth]{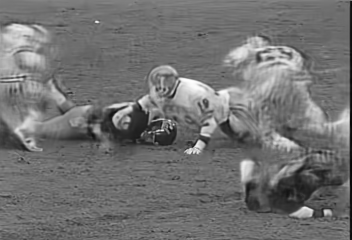}}  \vspace{-7pt}
\caption{Visual comparison of frames predicted by LFP-L2 vs. LFP-GAN models for the 9th frame of \textit{Football}.}
\label{fig:football}
\vspace{-3pt}
\end{figure}

\subsection{Frame Prediction Results}
\label{res:pred}
\vspace{-6pt}
We first provide quantitative and qualitative evaluation of our learned frame prediction~(LFP) models without considering the effect of compression; i.e., all results in this subsection use uncompressed frames as input. 

Evaluation of LFP-L1, LFP-L2, and LFP-GAN models vs. block motion-compensation (BMC) (using 16~$\times$~16 motion vectors and exhaustive search with 0.5 pixel accuracy) and frame difference~(FD) (with no prediction) is performed in terms of frame-by-frame PSNR of predicted videos plotted in Fig.~\ref{fig:prediction}. In these experiments, training samples are selected from entire UCF101 dataset with equal probability. The mean PSNR for each video (over all frames) as well as mean PSNR (over all videos) are tabulated in the first column of Table~\ref{tab:mean_psnr}. 

Sample frames from the sequence Football are shown in Fig.~\ref{fig:football} to demonstrate that predictions obtained by the LFP-GAN model are sharper, albeit having less fidelity to the original. More visual evaluation results can be accessed online at \url{https://serkansulun.com/lfp}

We conducted an additional experiment to evaluate the dependence of LFP-L2 model performance on the training set. To this effect, we sampled videos with type ``Sports", which have camera pan, with three times more probability. The results, in the second column of Table~\ref{tab:mean_psnr}, show around 1\% improvement in Garden and Mobile, which have camera pan, while the mean PSNR over all videos is almost unchanged. This result demonstrates that models can be trained for specific type of motion by choosing training samples with such motion.

Inspection of Table~\ref{tab:mean_psnr}, Fig.~\ref{fig:football}, and Fig.~\ref{fig:prediction} leads to the~following observations: \\
\indent 1) LFP-L2 vs. LPF-L1 model: The PSNR of LFP-L2 model exceeds  that of LFP-L1 model in videos with predictable low/moderate motion, such as Container and Hall Monitor; however, LFP-L1 model is slightly better for Harbour, Mobile, and Garden, which have more motion. This can be explained by that models trained by $\ell^1$ loss provide better generalization; hence, may perform slightly better for videos with less predictable motion.\\
\indent 2) LFP-L1/L2 vs. LPF-GAN model: The PSNR of LFP-L1 and LFP-L2 models exceeds that of LFP-GAN in all frames of all videos. The adversarial loss term that is added to the $\ell^2$ loss leads to visually more pleasing predicted frames 
at the expense of lower PSNR.\\
\indent 3) LFP-L1/L2 model vs. BMC: LFP-L1/L2 models outperform BMC for all frames in Harbour, Container, Garden, Hall  Monitor, and Mobile, which contain moderate predictable motion. LFP models are competitive with BMC in Coastguard and Tennis (except for frames where there is scene change). They are outperformed by BMC in Football and Foreman, which contain complex motions that are hard to predict. It is understandable that classic motion compensation, having access to the curent frame to be predicted and using motion vectors as side information, to perform better in such cases.
\vspace{-10pt}

\begin{table}[t]
\caption{Variation of prediction PSNR for LFP-L2 model according to selection of training sample video type.} \vspace{-14pt}
\begin{center}
\begin{tabular}{ccc}
       & Equiprobable & Sports more probable \\ \hline 
City         & 27.96    & 27.82      \\
Coastguard   & 32.00    & 32.07      \\
Container    & 41.50    & 41.37      \\
Football     & 22.87    & 22.77      \\
Foreman      & 31.38    & 31.34      \\
Garden       & 26.45    & 26.65      \\
Hall monitor & 36.60    & 36.57      \\
Harbour      & 28.38    & 28.26      \\
Mobile       & 27.31    & 27.70      \\
Tennis       & 29.88    & 29.57     \\ \hline
Mean PSNR    & 30.43    & 30.41      
\end{tabular}
\end{center}
\label{tab:mean_psnr}
\vspace{-20pt}
\end{table}

\subsection{Compression Efficiency Results}
\label{res:codec}
\vspace{-6pt}
We compare the rate-distortion (RD) performance of our LFP-L1-BPG, LFP-L2-BPG, and LFP-GAN-BPG methods with those of x264 \cite{x264} and x265 codecs~\cite{ffmpeg} configured for low-delay sequential (IPP...) coding in fixed QP, i.e.,~variable bitrate (VBR), setting. 
The performance of each codec is assessed based on RD (PSNR vs. bitrate) curves that are sampled at 11 bitrates corresponding to quantization parameters (QP) from 25 to 35 incremented by one for x264 codec, and from 20 to 30 for x265 codec (to approximately match the bitrates). 
The RD curves are compared using the Bjontegaard delta PSNR (BD-PSNR) metric \cite{bjontegaard}, which measures the~difference of areas between two RD curves. 

In order to analyze how much improvement comes from LFP vs. the BPG codec, we performed an ablation study. We first replaced BPG with WebP codec~\cite{webp} resulting in the LFP-L2-WEBP method to evaluate the~contribution~of the BPG codec. We next replaced LFP with $16~\times~16$ BMC using exhaustive search with 0.5 pixel accuracy and with FD with no prediction using the BPG codec in both cases. These methods are called BMC-BPG and FD-BPG, respectively, evaluates whether BPG alone is sufficient to get good results.

\begin{figure*}[h]
\centering
\subfloat      {
\includegraphics[width=0.80\linewidth]{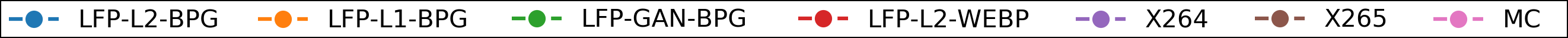}
} \vspace{-10pt} \\
\subfloat      {
\includegraphics[width=.342\linewidth]{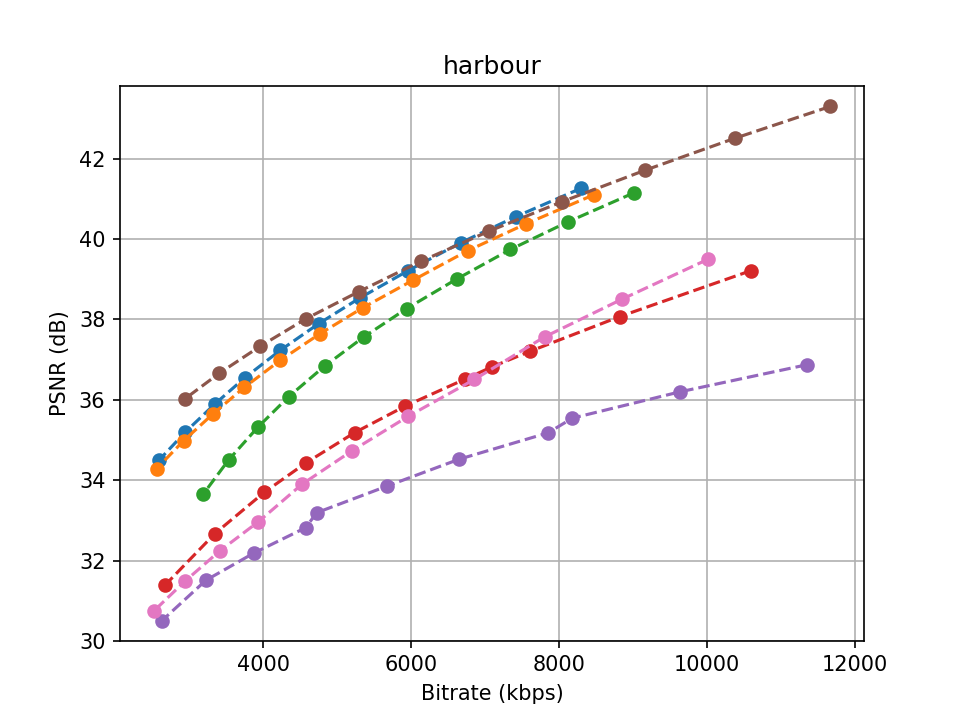}
} \hspace{-20pt}
\subfloat      {
\includegraphics[width=.342\linewidth]{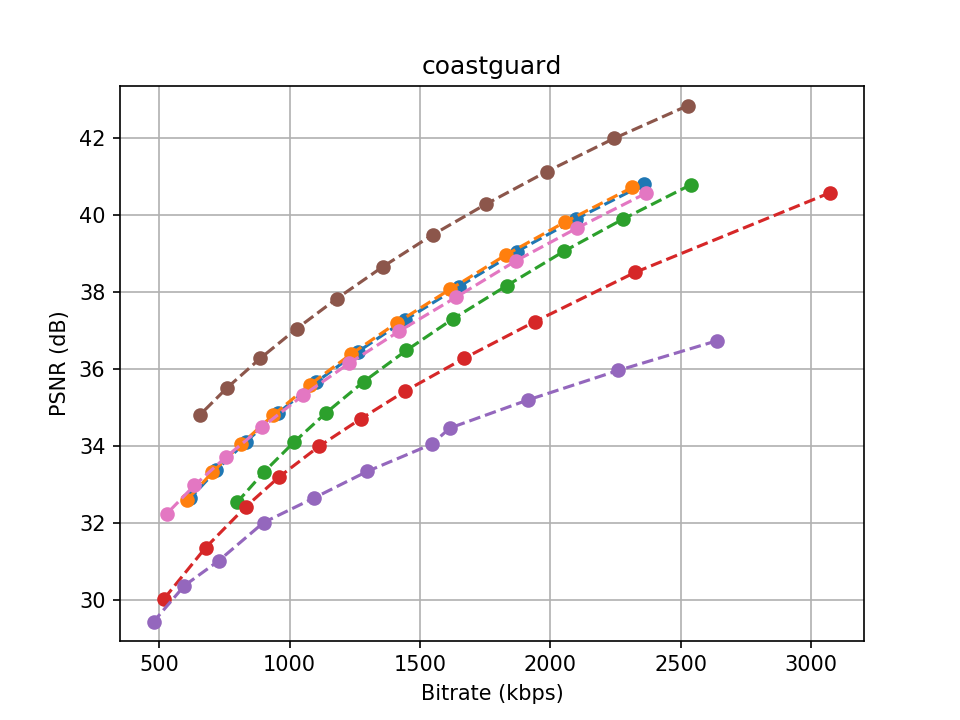}
} \hspace{-20pt} 
\subfloat     {
\includegraphics[width=.342\linewidth]{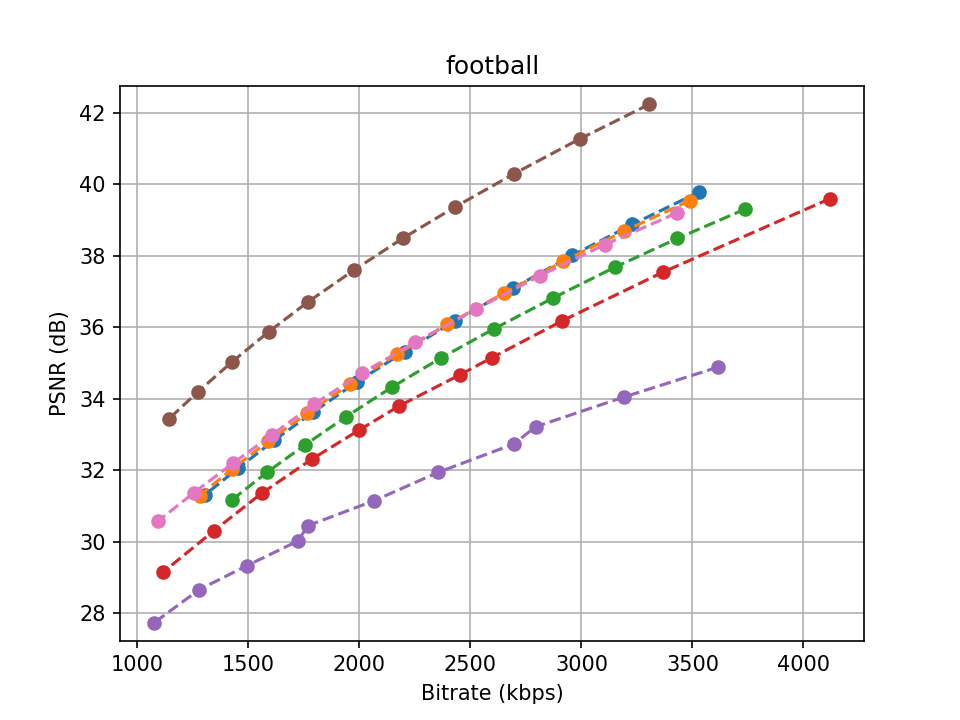}
} \vspace{-10pt} \\
\subfloat     {
\includegraphics[width=.342\linewidth]{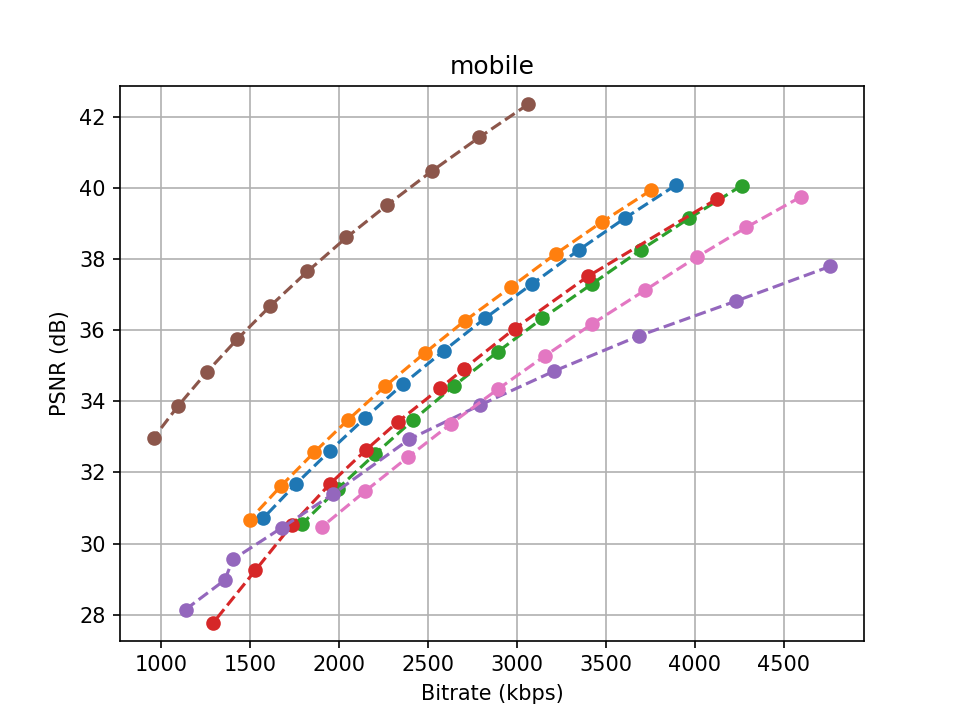}
} \hspace{-20pt} 
\subfloat     {
\includegraphics[width=.342\linewidth]{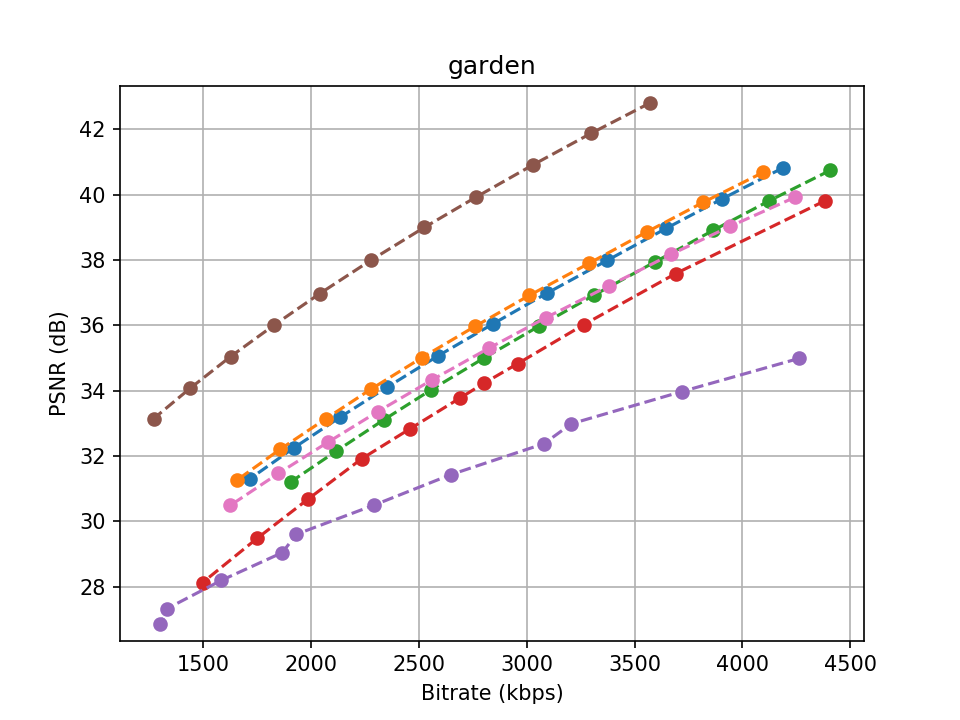}
} \hspace{-20pt}
\subfloat     {
\includegraphics[width=.342\linewidth]{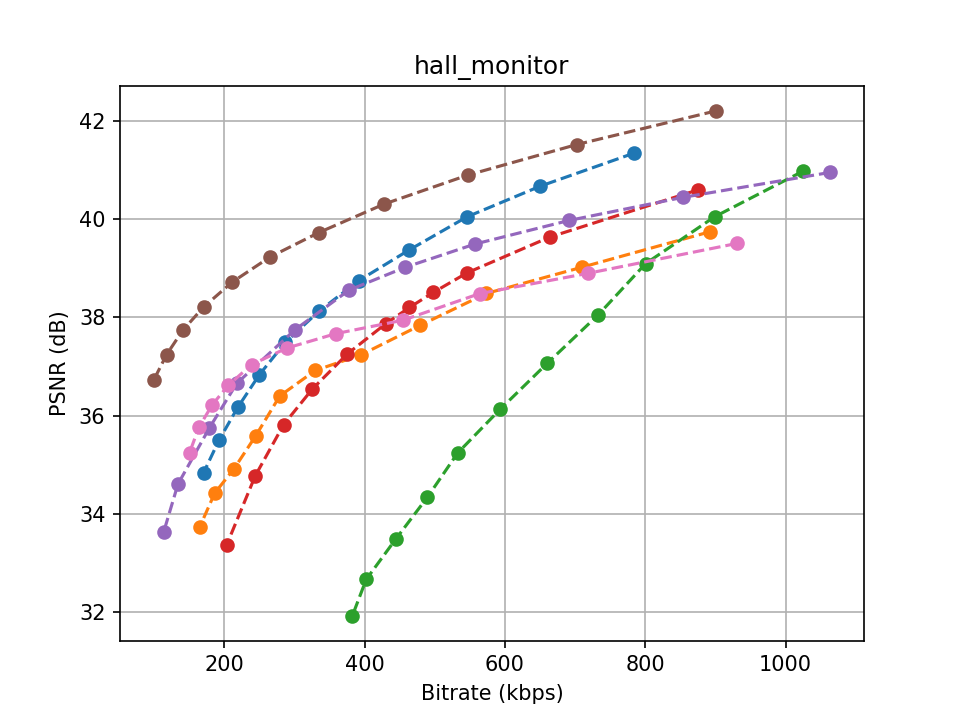}
} \vspace{-10pt} \\
\subfloat     {
\includegraphics[width=.342\linewidth]{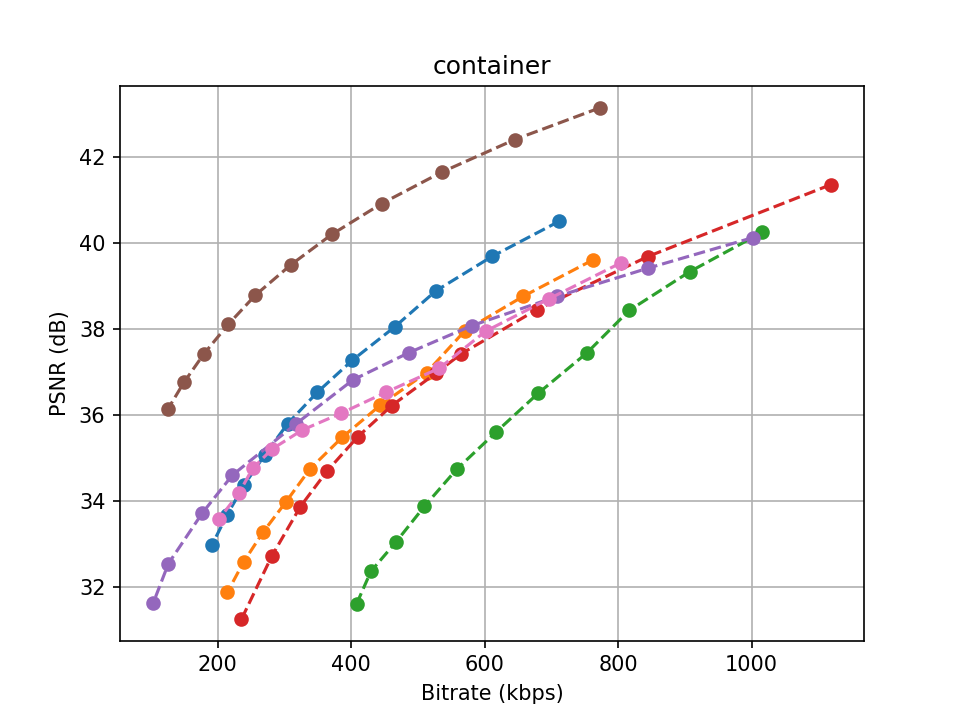}
} \hspace{-20pt}
\subfloat     {
\includegraphics[width=.342\linewidth]{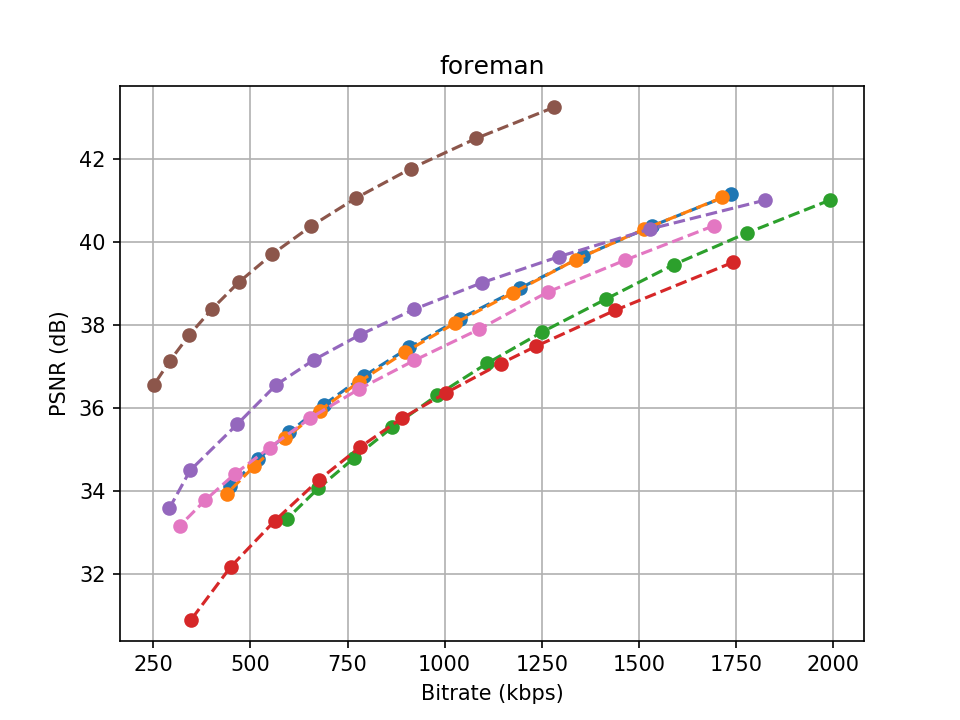}
} \hspace{-20pt} 
\subfloat     {
\includegraphics[width=.342\linewidth]{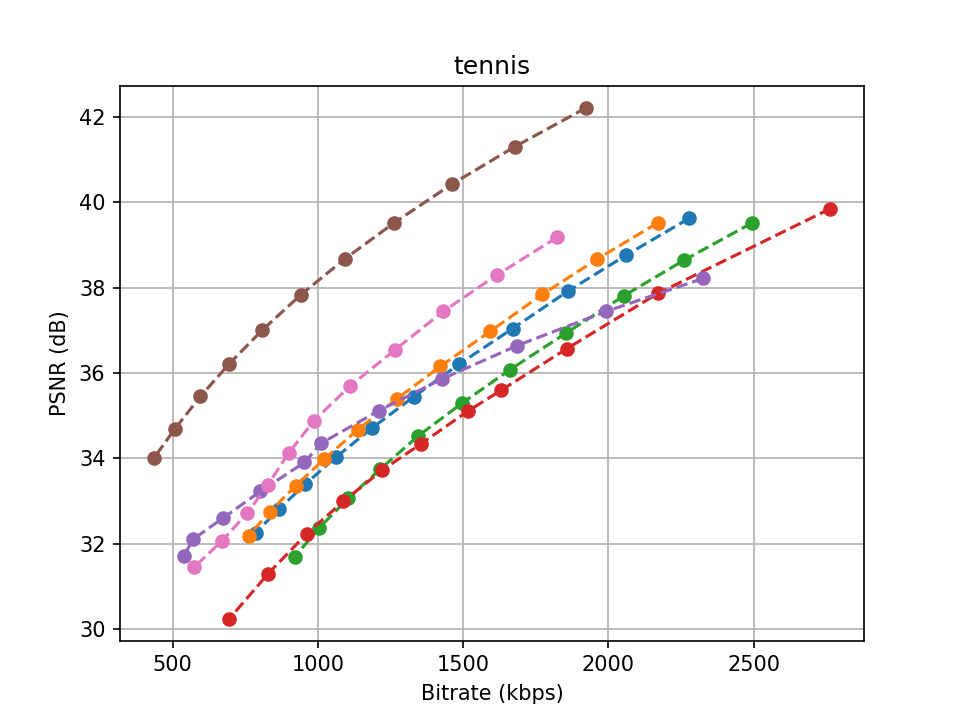}
} 
\vspace{-2pt} \\
\caption{Comparison of compression efficiency of different LFP models versus standards-based video codecs: PSNR~vs.~bitrate curves for LFP-L2-BPG, LFP-L1-BPG, LFP-GAN-BPG, LFP-L2-WebP, BMC-BPG, x264, and x265 codecs (in sequential low-delay mode) on nine MPEG test sequences.
}
\label{fig:compression}
\vspace{12pt}
\end{figure*}

The~PSNR vs. bitrate curves for all test sequences are depicted in Fig.~\ref{fig:compression}. Table~\ref{table:bjontegaard} shows the Bjontegaard delta PSNR (BD-PSNR) with respect to the anchor x264 sequential averaged over 10 videos, which reveals: \vspace{1pt} \\
\indent 1) LFP-L2-BPG, LFP-L1-BPG and LFP-GAN-BPG methods outperform the anchor x264 sequential on average by 1.779, 1.524 and 0.57 dB, respectively. \vspace{1pt} \\
\indent 2) LFP-L2-WEBP method, where the RD performance of WebP codec is similar to that of x264 intra mode, also outperforms x264 sequential on average by 0.645 dB showing the power of LFP-L2 model. \vspace{1pt} \\
\indent 3) The average BD-PSNR of BMC-BPG method is 0.856 dB, which is lower than that of LFP-L2-BPG, again showing the power of LFP-L2 model.
\vspace{1pt} \\
\indent 4) As shown by the output samples, even though the LFP-GAN model generates sharper and visually more pleasing predicted images, the LFP-GAN-BPG method is inferior to LFP-L1/L2-BPG in terms of BD-PSNR. \vspace{1pt} \\
\indent 5) The LFP-L2-BPG method approaches the performance of x265 (in low-delay mode) only for Harbour. 

\begin{figure*}
\centering
\subfloat      {
\includegraphics[width=0.6\linewidth]{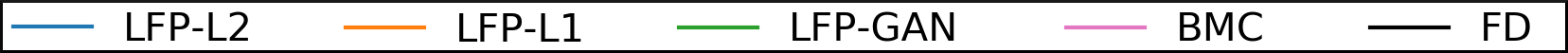}
} \vspace{-10pt} \\
\subfloat     {
\includegraphics[width=.34\linewidth]{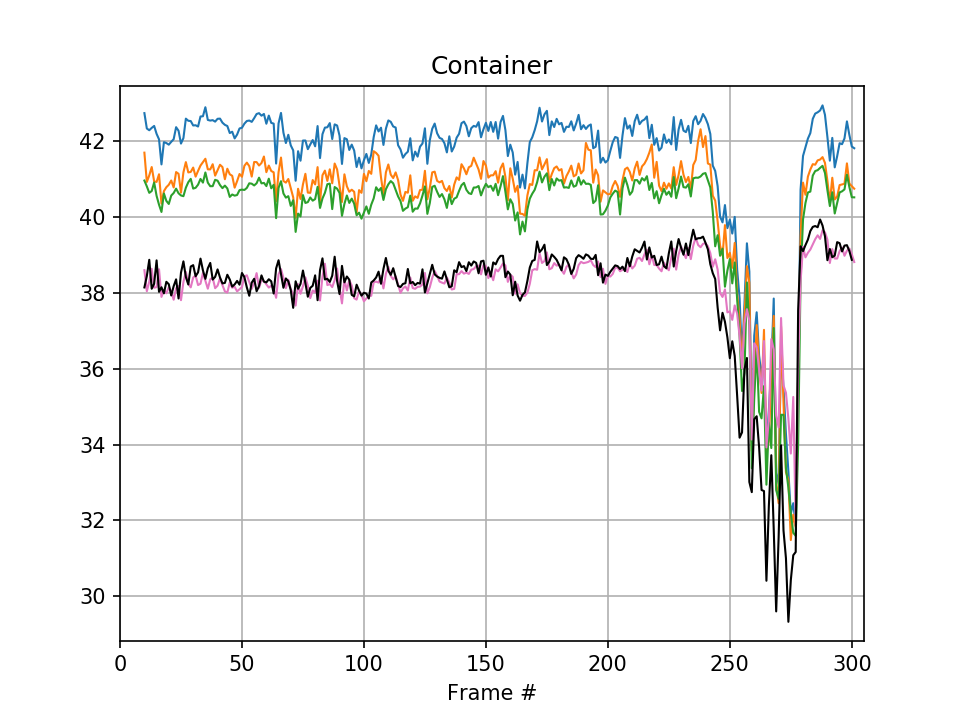}
} \hspace{-22pt}
\subfloat     {
\includegraphics[width=.34\linewidth]{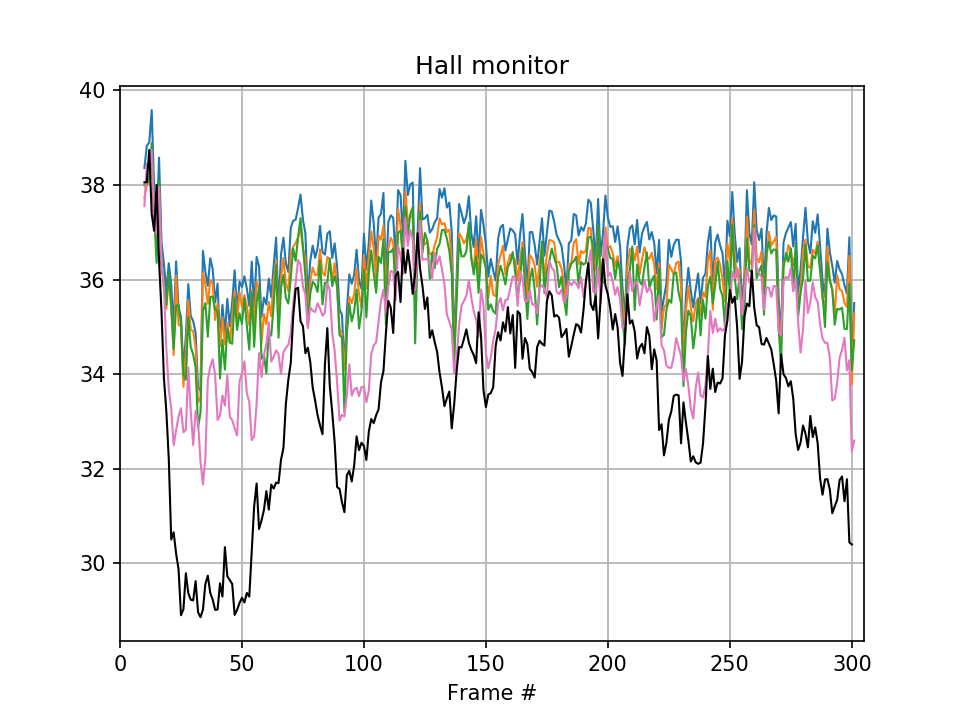}
} \hspace{-22pt}
\subfloat     {
\includegraphics[width=.34\linewidth]{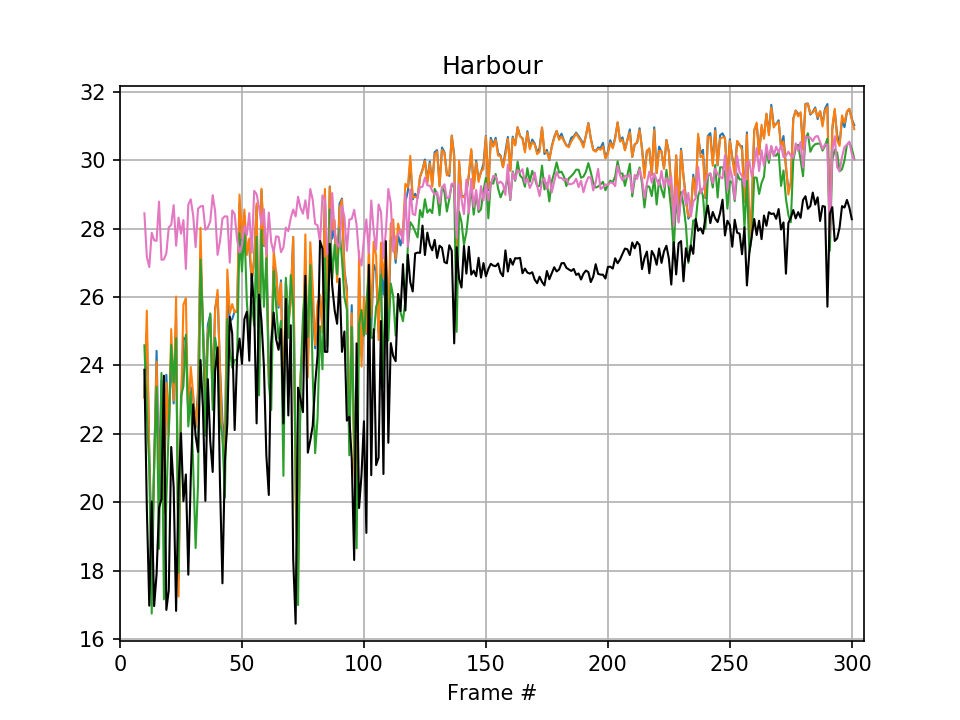}
} \vspace{-15pt} \\
\subfloat     {
\includegraphics[width=.34\linewidth]{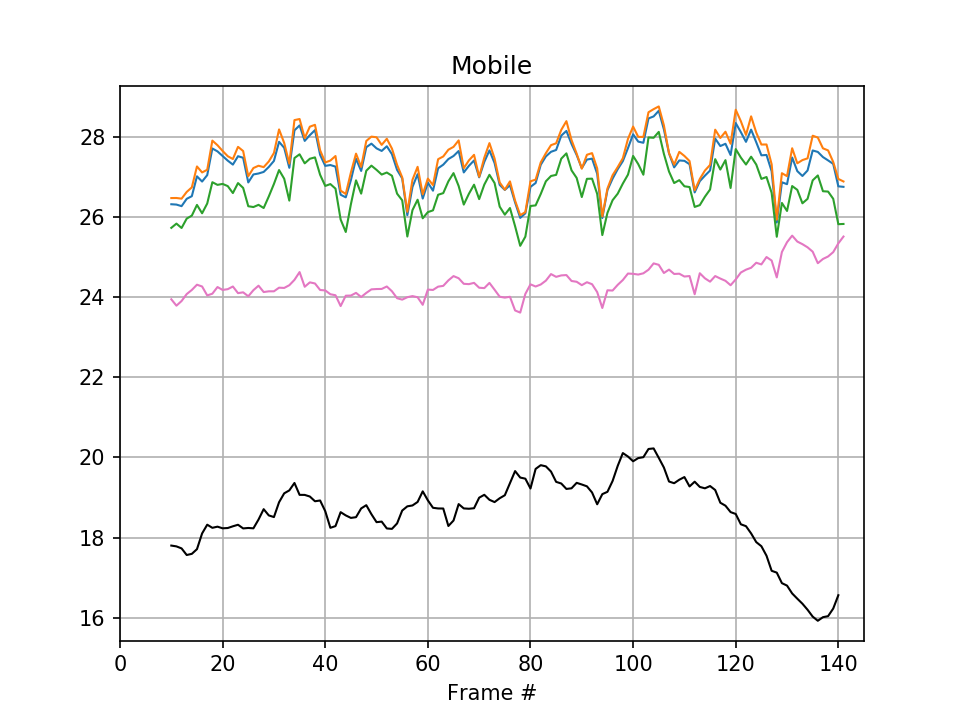}
} \hspace{-22pt}
\subfloat     {
\includegraphics[width=.34\linewidth]{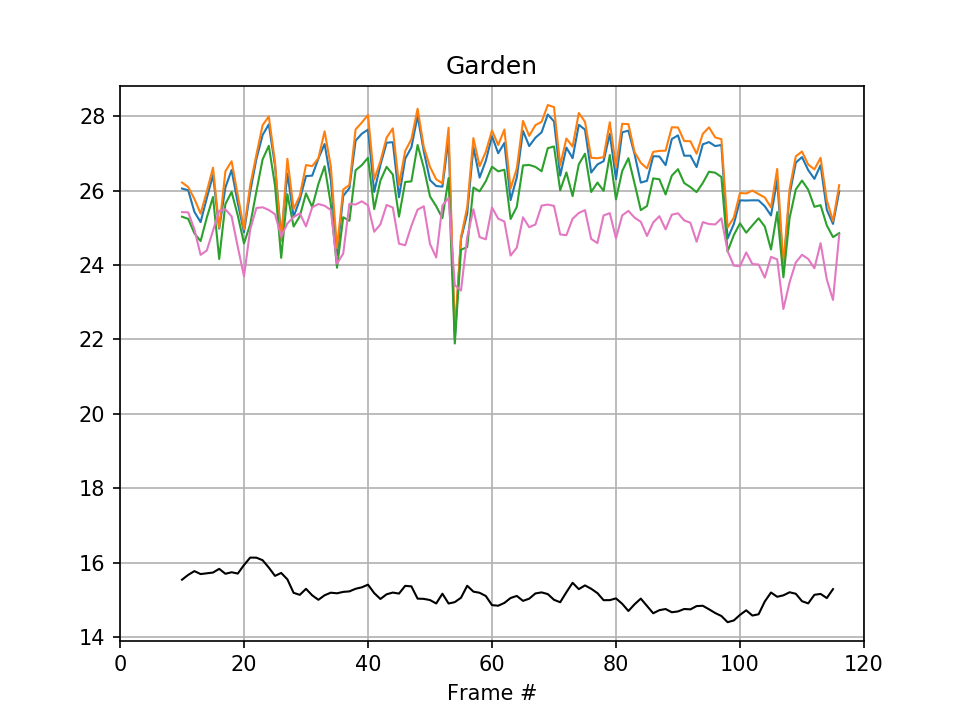}
} \hspace{-22pt}
\subfloat     {
\includegraphics[width=.34\linewidth]{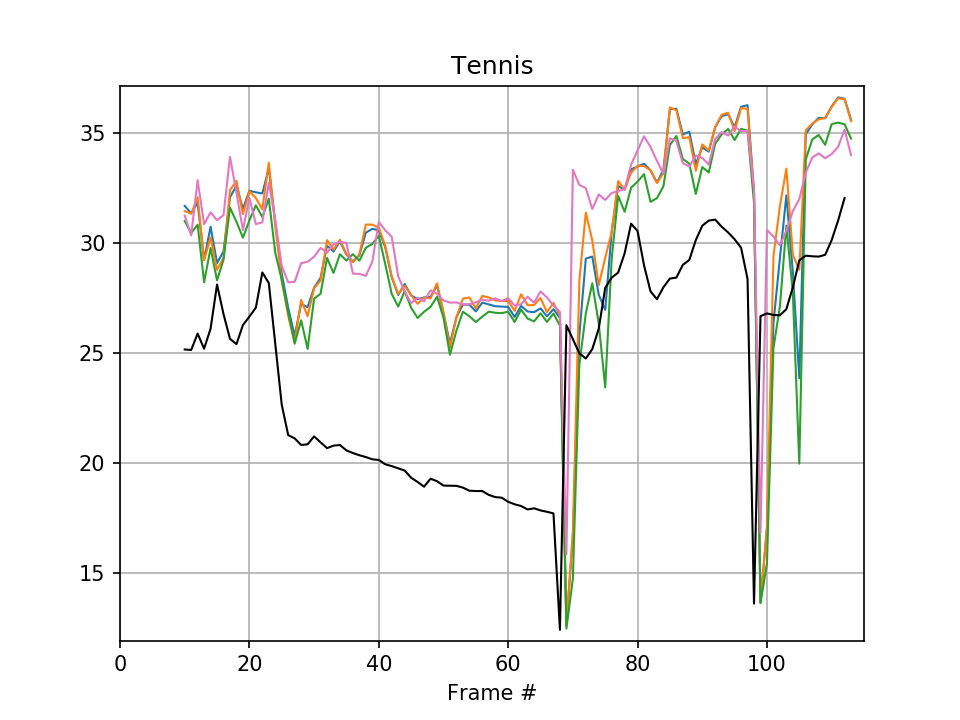}
} \vspace{-15pt} \\
\subfloat      {
\includegraphics[width=.34\linewidth]{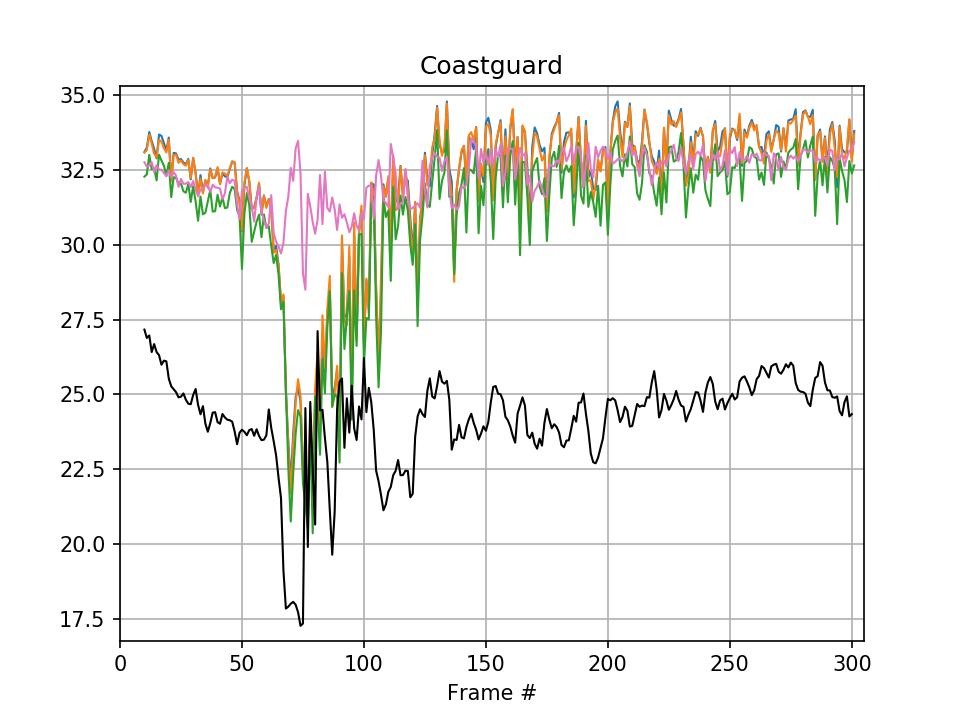}
} \hspace{-22pt}
\subfloat     {
\includegraphics[width=.34\linewidth]{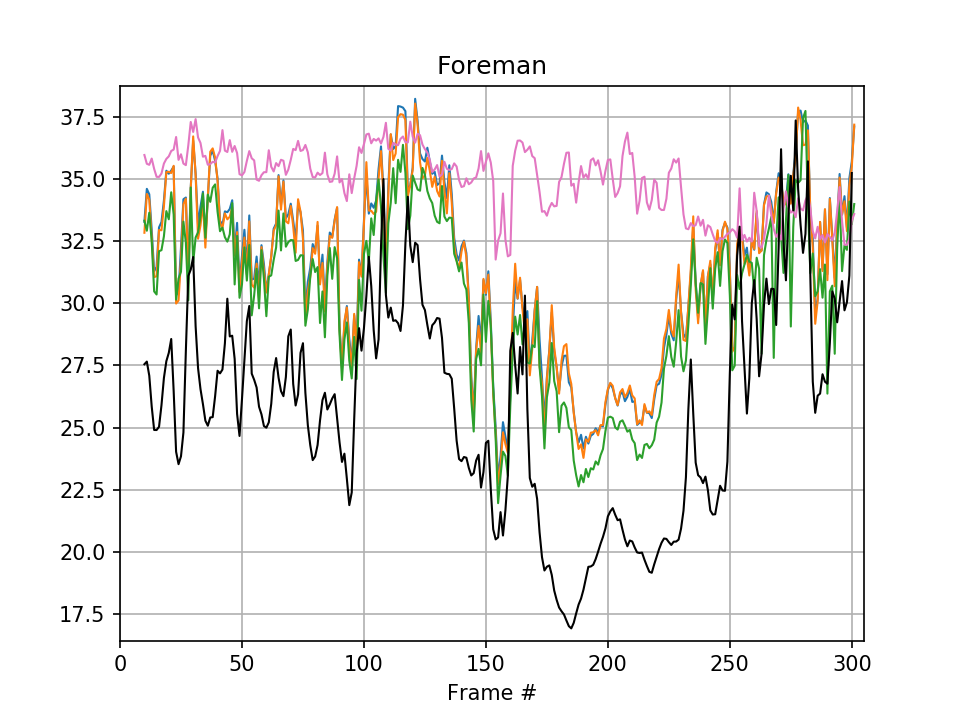}
} \hspace{-22pt}
\subfloat     {
\includegraphics[width=.34\linewidth]{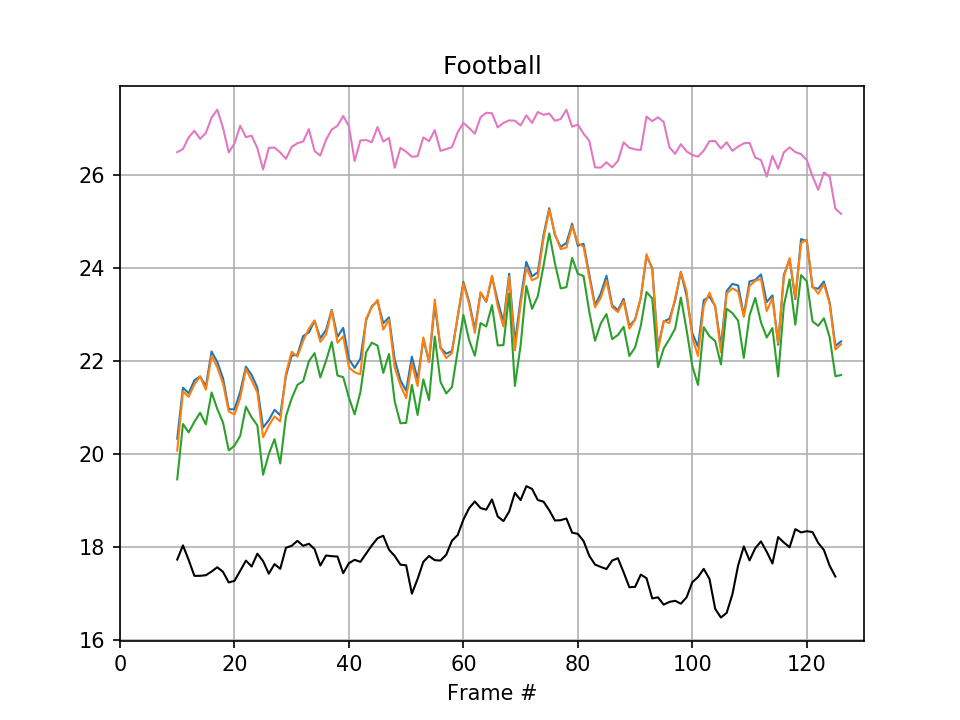}
} 
\caption{Comparison of frame prediction performance: PSNR of predicted frames vs. frame number to compare performances of learned frame prediction (LFP) models trained by mean square loss (LFP-L2), by $\ell^1$ loss (LFP-L1), and by a combination of mean square loss and adversarial loss~(LFP-GAN) vs. 16x16 block motion compensation with 0.5 pixel accuracy using exhaustive search (BMC) and frame difference with no prediction (FD) for all test videos. The LFP-L2 and LFP-L1 models trained on UCF101 dataset perform the best for all frames on Harbour, Container, Hall monitor, Mobile, and Garden videos. They are competitive with BMC in Tennis and Coastguard.
}
\label{fig:prediction}
\end{figure*}

Analyzing the RD performance of the LFP-L2-BPG method on individual videos, we observe that it outperforms all other methods, except x265, even in the case of high motion videos, such as Football, where the frame prediction performance of LFP-L2 model for Football was inferior to BMC according to Fig.~\ref{fig:prediction}. 

Several factors may lead to this result: i)~Motion vector overhead of high motion videos, such as Football, is significant, especially at low bitrates,
while the LFP-BPG method has none; hence, encodes residuals better.
ii) Prediction based on compressed-decompressed frames affects BMC more adversely than it affects LFP.

\section{Conclusion}
\label{conclusion}
\vspace{-10pt}
We demonstrate that the average RD performance over a diverse set of MPEG test videos of a simple video codec using a universal LFP model trained with $\ell^2$ loss and an open source intra (BPG) codec exceeds that of the well-established x264 codec using variable size block motion compensation. 
The proposed approach achieves surprisingly good results in videos with predictable motion, such as Harbour. We reach the following conclusions:
1)~The proposed LFP method can predict continuous motions, including camera pan, effectively.
2)~Videos with complex motion are harder to predict. It is understandable that standards-based motion compensation, having access to the current frame to be predicted and sending motion vectors as side information, to perform better prediction in such cases. Yet, it is interesting to see that compression performance of LFP+BPG is still competitive with that of x264 for Football, because LFP+BPG does not require sending side information and can use those bits to encode the residual better. The~compression efficiency of x265 codec (in both prediction as well as coding residual and side information) is clearly better than that of LFP+BPG.
3)~Videos with complex motion exhibit a large variation (e.g., occlusions, scene transitions, etc.) that is very difficult to model. One cannot guarantee that training a network with a class of complex motion videos will generalize well to test videos with different type of complex motion.


\clearpage

\begin{table*}[h]
\centering
\caption{Bjontegaard delta PSNR (BD-PSNR) versus x264 (anchor). Positive values show better performance.}
\begin{tabular}{c|ccccccc}
             & \shortstack{LFP \\ L2 \\ BPG} & \shortstack{LFP \\ L1 \\ BPG} & \shortstack{LFP \\ GAN \\ BPG} & \shortstack{LFP \\ L2 \\ WEBP} & X265  & \shortstack{ BMC \\ BPG}   & \shortstack{ FD \\ BPG}    \\ \hline
Harbour      & 4.801      & 4.560      & 3.895       & 2.338       & 5.354 & 1.355  & -0.557 \\
Garden       & 3.998      & 4.129      & 3.399       & 2.252       & 7.461 & 3.283  & -7.025 \\
Football     & 3.738      & 3.742      & 3.078       & 2.456       & 6.658 & 3.632  & -2.472 \\
Coastguard   & 3.193      & 3.219      & 2.597       & 1.900       & 5.056 & 3.027  & -6.240 \\
Mobile       & 2.017      & 2.239      & 1.194       & 1.028       & 6.637 & 0.441  & -4.776 \\
Container    & 0.382      & -1.017     & -2.410      & -0.115      & 3.755 & -0.382 & -2.101 \\
City         & 0.283      & 0.282      & -0.224      & -1.483      & 3.205 & -1.838 & -4.333 \\
Hall monitor & 0.116      & -1.259     & -3.140      & 0.338       & 2.070 & -0.556 & -2.753 \\
Tennis       & 0.064      & 0.208      & -0.690      & -0.638      & 3.866 & 0.680  & -4.420 \\
Foreman      & -0.799     & -0.864     & -1.999      & -1.626      & 3.389 & -1.083 & -6.530 \\ \hline
AVERAGE      & 1.779      & 1.524      & 0.570       & 0.645       & 4.745 & 0.856  & -4.121
\end{tabular}
\label{table:bjontegaard}
\end{table*}


\bibliography{references}

\begin{thebibliography}{10}
\providecommand{\url}[1]{{#1}}
\providecommand{\urlprefix}{URL }
\expandafter\ifx\csname urlstyle\endcsname\relax
  \providecommand{\doi}[1]{DOI~\discretionary{}{}{}#1}\else
  \providecommand{\doi}{DOI~\discretionary{}{}{}\begingroup
  \urlstyle{rm}\Url}\fi

\bibitem{webp}
A new image format for the web.
\newblock \urlprefix\url{https://developers.google.com/speed/webp}

\bibitem{x264}
x264: A high performance h.264/avc encoder.
\newblock https://www.videolan.org/developers/x264.html  (2006)

\bibitem{babaeizadeh2017stochastic}
Babaeizadeh, M., Finn, C., Erhan, D., Campbell, R.H., Levine, S.: Stochastic
  variational video prediction.
\newblock In: Int. Conf. Learn. Rep. (ICLR) Vancouver, Canada (2018)

\bibitem{bai2018}
Bai, S., Kolter, J.Z., Koltun, V.: An empirical evaluation of generic
  convolutional and recurrent networks for sequence modeling.
\newblock arXiv preprint https://arxiv.org/pdf/1803.01271.pdf  (2018)

\bibitem{bpg}
Bellard, F.: Better portable graphics.
\newblock https://www.bellard.org/bpg [Last accessed: Apr.~2020]

\bibitem{ffmpeg}
Bellard, F.: Ffmpeg multimedia system.
\newblock https://www.ffmpeg.org/ [Last accessed: Apr. 2020]

\bibitem{bjontegaard}
Bjontegaard, G.: Calculation of average psnr differences between rd-curves.
\newblock VCEG-M33  (2001)

\bibitem{zchen2018}
Chen, Z., He, T., Jin, X., Wu, F.: Learning for video compression.
\newblock IEEE Trans. on Circuits and Systems for Video Technology
  \textbf{30}(2), 566--576 (2020)

\bibitem{ganhacks}
Chintala, S., Denton, E., Arjovsky, M., Mathieu, M.: How to train a {GAN}?
  {T}ips and tricks to make {GANs} work (2016).
\newblock \urlprefix\url{https://github.com/soumith/ganhacks}

\bibitem{cho2019}
Choi, H., Bajic, I.V.: Deep frame prediction for video coding.
\newblock IEEE Trans. Circ. Syst. Video Tech.  (2019)

\bibitem{chollet}
Chollet, F.: Deep {Learning} with {Python}.
\newblock Manning Publications Company (2017)

\bibitem{denton2018stochastic}
Denton, E., Fergus, R.: Stochastic video generation with a learned prior.
\newblock In: Proc. of Int. Conf. on Machine Learning (PMLR), pp. 80:1174--1183
  (2018)

\bibitem{perceptualloss}
Dosovitskiy, A., Brox, T.: Generating images with perceptual similarity metrics
  based on deep networks.
\newblock In: Adv. in Neural Infor. Proc. Systems, pp. 658--666 (2016)

\bibitem{dumas2018}
Dumas, T., Roumy, A., Guillemot, C.: Autoencoder based image compression: {Can}
  the learning be quantization independent?
\newblock In: IEEE ICASSP, Calgary, Canada (2018)

\bibitem{finn}
Finn, C., Goodfellow, I., Levine, S.: Unsupervised learning for physical
  interaction through video prediction.
\newblock In: Adv. in Neural Infor. Proc. Systems, pp. 64--72 (2016)

\bibitem{iscas2018}
Huo, S., Liu, D., Wu, F., Li, H.: Convolutional neural network-based motion
  compensation refinement for video coding.
\newblock In: IEEE Int. Symp. on Circuits and Systems (ISCAS), Florence, Italy
  (May 2018)

\bibitem{kalchbrenner}
Kalchbrenner, N., Oord, A.v.d., Simonyan, K., Danihelka, I., Vinyals, O.,
  Graves, A., Kavukcuoglu, K.: Video pixel networks.
\newblock In: Proc. of Int. Conf. on Machine Learning (PMLR), pp. 70:1771--1779
  (2017)

\bibitem{adam}
Kingma, D.P., Ba, J.: Adam: A method for stochastic optimization.
\newblock In: Int. Conf. on Rep. Learn. (ICLR) (2015)

\bibitem{lee2018stochastic}
Lee, A.X., Zhang, R., Ebert, F., Abbeel, P., Finn, C., Levine, S.: Stochastic
  adversarial video prediction.
\newblock arXiv:1804.01523  (2018)

\bibitem{edsr}
Lim, B., Son, S., Kim, H., Nah, S., Lee, K.M.: Enhanced deep residual networks
  for single image super-resolution.
\newblock In: IEEE Conf. on Computer Vision and Pattern Recognition Workshops
  (CVPRW), vol.~1, p.~4 (2017)

\bibitem{lin2018}
Lin, J., Liu, D., Li, H., Wu, F.: Generative adversarial network-based frame
  extrapolation for video coding.
\newblock In: Visual Comm. and Image Proc. (VCIP) (2018)

\bibitem{lu2018}
Lu, G., Zhang, X., Chen, L., Gao, Z.: Novel integration of frame rate up
  conversion and {HEVC} coding based on rate-distortion optimization.
\newblock IEEE Trans. on Image Proc. \textbf{27}(2), 678--691 (2018)

\bibitem{mathieu}
Mathieu, M., Couprie, C., LeCun, Y.: Deep multi-scale video prediction beyond
  mean square error.
\newblock In: Proc. of Int. Conf. on Learning Representation (ICLR) (2016)

\bibitem{pixelcnn}
Oord, A.v.d., Kalchbrenner, N., Kavukcuoglu, K.: Pixel recurrent neural
  networks.
\newblock In: Proc. of Int. Conf. on Machine Learning (ICML), vol.~48, pp.
  1747--–1756 (2016)

\bibitem{dcgan}
Radford, A., Metz, L., Chintala, S.: Unsupervised representation learning with
  deep convolutional generative adversarial networks.
\newblock In: ICLR (Poster) (2016)

\bibitem{xia2019}
S.~Xia W.~Yang, Y.H., Liu, J.: Deep inter prediction via pixel-wise motion
  oriented reference generation.
\newblock In: IEEE Int. Conf. Image Proc. (2019)

\bibitem{heiko2016}
Schwarz, H., Wiegand, T.: Video coding: {Part II} of fundamentals of source and
  video coding.
\newblock Foundations and Trends in Signal Processing \textbf{10}(1-3), 1--346
  (2016)

\bibitem{survey}
Selva~Castell{\'o}, J.: A comprehensive survey on deep future frame video
  prediction.
\newblock Master's thesis, Universitat Polit{\`e}cnica de Catalunya (2018)

\bibitem{pixelshuffler}
Shi, W., et~al.: Real-time single image and video super-resolution using an
  efficient sub-pixel convolutional neural network.
\newblock In: IEEE Conf. CVPR, pp. 1874--1883 (2016)

\bibitem{ucf101}
Soomro, K., Zamir, A.R., Shah, M.: Ucf101: A dataset of 101 human actions
  classes from videos in the wild.
\newblock arXiv:1212.0402  (2012)

\bibitem{srivastava}
Srivastava, N., Mansimov, E., Salakhudinov, R.: Unsupervised learning of video
  representations using {LSTMs}.
\newblock In: Int. Conf. on Machine Learning, pp. 843--852 (2015)

\bibitem{residualscaling}
Szegedy, C., Ioffe, S., Vanhoucke, V., Alemi, A.A.: Inception-v4,
  inception-resnet and the impact of residual connections on learning.
\newblock In: AAAI, vol.~4, p.~12 (2017)

\bibitem{ntire}
Timofte, R., et~al.: {NTIRE} 2017 challenge on single image super-resolution:
  Methods and results.
\newblock In: IEEE Conf. Computer Vision and Pattern Recognition Workshops
  (CVPRW), pp. 1110--1121 (2017)

\bibitem{ntire2018}
Timofte, R., et~al.: {NTIRE} 2018 challenge on single image super-resolution:
  Methods and results.
\newblock In: IEEE Conf. Computer Vision and Pattern Recognition Workshops
  (CVPRW), pp. 965--976 (2018)

\bibitem{amersfoort}
Van~Amersfoort, J., Kannan, A., Ranzato, M., Szlam, A., Tran, D., Chintala, S.:
  Transformation-based models of video sequences.
\newblock arXiv:1701.08435  (2017)

\bibitem{nips2019}
Villegas, R., Pathak, A., Kannan, H., Erhan, D., Le, Q.V., Lee, H.: High
  fidelity video prediction with large stochastic recurrent neural networks.
\newblock In: Conf. on Neural Information Processing Systems (NIPS) (2019)

\bibitem{villegaslongterm}
Villegas, R., Yang, J., Zou, Y., Sohn, S., Lin, X., Lee, H.: Learning to
  generate long-term future via hierarchical prediction.
\newblock In: Int. Conf. Mach. Learning (ICML) (2017)

\bibitem{vondrick}
Vondrick, C., Torralba, A.: Generating the future with adversarial
  transformers.
\newblock In: IEEE Conf. Computer Vision and Pattern Recog. (CVPR), vol.~1,
  p.~3 (2017)

\bibitem{esrgan}
Wang, X., et~al.: {ESRGAN}: {Enhanced} super-resolution generative adversarial
  networks.
\newblock In: Proc. of the {European} {Conf.} on {Computer} {Vision} ({ECCV})
  (2018)

\bibitem{wang2018}
Wang, Y., Fan, X., Jia, C., Zhao, D., Gao, W.: Neural network based inter
  prediction for {HEVC}.
\newblock In: IEEE Int. Conf. Multimedia and Expo (2018)

\bibitem{wichers2018hierarchical}
Wichers, N., Villegas, R., Erhan, D., Lee, H.: Hierarchical long-term video
  prediction without supervision.
\newblock In: Proc. of Int. Conf. on Mach. Learn. (PMLR), Stockholm (2018)

\bibitem{zhao2018}
Zhao, L., Wang, S., Zhang, X., Wang, S., Ma, S., Gao, W.: Enhanced {CTU}-level
  inter prediction with deep frame rate up-conversion for high efficiency video
  coding.
\newblock In: IEEE Int. Conf. Image Proc. (2018)

\end{thebibliography}

%
%


%
%

\bibliographystyle{spmpsci}      

%
%

\end{document}